\documentstyle[eqsecnum,aps,prc,floats,graphicx,tighten]{revtex}

\def\today{\space\number\day\space\ifcase\month\or January\or February\or
   March\or April\or May\or June\or July\or August\or September\or
   October\or November\or December\fi\space\number\year}

\begin{document}
\draft  

\twocolumn[\hsize\textwidth\columnwidth\hsize\csname@twocolumnfalse\endcsname

\title{New Limit on the $D$ Coefficient in Polarized Neutron Decay}

\author{L.\,J.\,Lising$^{1}$\footnotemark, S.\,R.\,Hwang$^{2}$\footnotemark,
J.\,M.\,Adams$^{3}$,
T.\,J.\,Bowles$^{4}$, M.\,C.\,Browne$^{4}$, T.\,E.\,Chupp$^{2}$, K.\,P.\,Coulter$^{2}$,
M.\,S.\,Dewey$^{3}$,
S.\,J.\,Freedman$^{1}$, B.\,K.\,Fujikawa$^{1}$, A.\,Garcia$^{5}$,
G.\,L.\,Greene$^{4}$,
G.\,L.\,Jones$^{3}$\footnotemark, H.\,P.\,Mumm$^{7}$, J.\,S.\,Nico$^{3}$,
J.\,M.\,Richardson$^{6}$\footnotemark, R.\,G.\,H.\,Robertson$^{7}$,
T.\,D.\,Steiger$^{7}$, W.\,A.\,Teasdale$^{4}$,
A.\,K.\,Thompson$^{3}$, E.\,G.\,Wasserman$^{1}$\footnotemark,
F.\,E.\,Wietfeldt$^{3}$, R.\,C.\,Welsh$^{2}$,
J.\,F.\,Wilkerson$^{7}$}
\address{\it $^{1}$ University of California and Lawrence Berkeley
National Laboratory,
Berkeley, California 94720}
\address{\it $^{2}$ University of Michigan, Ann Arbor, Michigan 48101}
\address{\it  $^{3}$ National Institute of Standards and Technology,
Gaithersburg, Maryland 20899}
\address{\it $^{4}$ Los Alamos National Laboratory, Los Alamos, New Mexico
87545}
\address{\it $^{5}$ University of Notre Dame, Notre Dame, Indiana
46556}
\address{\it $^{6}$ Harvard University, Cambridge, Massachusetts 02138}
\address{\it $^{7}$ University of Washington, Seattle, WA 98195}

\author{(The emiT Collaboration)}

\date{\today}
\maketitle

\begin{abstract}

We describe an experiment that has set new limits on the time reversal
invariance violating $D$ coefficient in neutron beta-decay. The emiT experiment
measured the angular correlation
$\langle \mathbf{J}\rangle \cdot({\mathbf p}_{e}\times {\mathbf p}_{p})$
using an octagonal symmetry that optimizes electron-proton coincidence rates.
The result is $D= \lbrack -0.6\pm 1.2({\rm stat}) \pm 0.5({\rm syst})
\rbrack \times 10^{-3}$. This
improves constraints
on the phase of $g_A/g_V$ and limits contributions to  $T$ violation
due to leptoquarks.
This paper
 presents details of the experiment, data analysis, and the investigation of
 systematic effects.

\end{abstract}
\pacs{PACS number(s): 11.30.Er, 12.15.Ji, 13.30.Ce, 23.20 En }

] 

\section{Introduction}\label{intro}

 $CP$ violation has been observed so far only in the decays of neutral
kaons \cite{KCP}.  Recently evidence for the implied $T$ violation in
the neutral kaon system has  been reported \cite{CPLear}.
These effects could be due to the Kobayashi-Maskawa phase in the
Standard Model \cite{Koba73}.  However, these observations could also be
due to
new physics, and it is well-established that new sources of $CP$
violation are required by the
observed baryon asymmetry of the universe \cite{BAUrefs}.
Many extensions of the  Standard
 Model contain new
sources of $CP$ violation and can be probed in
observables for which the contribution of the Kobayashi-Maskawa phase
in the Standard Model is small.   The
present experiment searches for $CP$ violation in one such observable, a
$T$-odd
correlation in  the decay of free
neutrons.

The differential decay rate for a free neutron can be
written \cite{Jack57}
\begin{eqnarray}
dW\propto
S(E_{e})dE_e d\Omega_{e} d\Omega_{\nu}
\left[
1+a\frac{\mathbf{p}_{e}\cdot\mathbf{p}_{\nu}}{E_{e}E_{\nu}}
\right.
\nonumber \\
\left.
 +\frac{\langle\mathbf{J}\rangle}{J}\cdot\left(
A\frac{\mathbf{p}_{e}}{E_e}+B\frac{\mathbf{p}_{\nu}}{E_{\nu}} +
D\frac{\mathbf{p}_{e}\times\mathbf{p}_{\nu}}{E_{e}E_{\nu}}
\right)
\right],
\label{dW(ps)}
\end{eqnarray}
where $p_{e}$, $E_{e}$ and $p_{\nu}$, $E_{\nu}$
are the momentum and energy of the outgoing electron and neutrino,
respectively,
$S(E_{e})$ is a phase space factor,
and $\langle\mathbf{J}\rangle$ is  the neutron spin.
The triple-correlation
$D\langle\mathbf{J}\rangle\cdot(\mathbf{p}_{e}\times\mathbf{p}_{\nu})$ is
odd under motion reversal, and can be used to measure
time reversal invariance violation when final state interactions are
taken into account.
Note that in the rest frame of the neutron, conservation of
momentum allows the transformation of the triple-correlation term
into
\begin{displaymath}
-D\frac{\langle\mathbf{J}\rangle}{J}\cdot
\frac{\mathbf{p}_{e}\times\mathbf{p}_{p}}{E_{e}E_{\nu}}
\end{displaymath}
where $\mathbf{p}_{p}$ is the momentum of the recoil proton.

  \footnotetext[1]{Present address:  National Institute of Standards and
Technology, Gaithersburg, Maryland}
  \footnotetext[2]{Present address:  National Central University,
  Chung-Li, Taiwan}
  \footnotetext[3]{Present address:  Hamilton College, Clinton, NY}
  \footnotetext[4]{Present address:  SAIC, Somerville, MA}
  \footnotetext[5]{Present address:  Personify, Inc., San Francisco, CA}

The $D$ coefficient is sensitive only to $T$-odd interactions with
vector and axial
vector
currents.  In a theory with such currents, the coefficients of the
correlations depend on the
magnitude and phase of $\lambda =|\lambda| e^{-i\phi}$, where
$|\lambda| = |g_A/g_V|$ is the magnitude the ratio of the axial vector to
vector
form factors of the nucleon.   In this notation, the coefficients are
given by
\begin{eqnarray}
a= \frac {1 - |\lambda|^2}{1 +3 |\lambda|^2},
\quad A= -2\frac {|\lambda| \cos \phi + |\lambda|^2}{1 +3 |\lambda|^2},
\nonumber \\
  B= -2\frac {|\lambda| \cos \phi - |\lambda|^2}{1 +3 |\lambda|^2},
  \quad D=
2\frac{|\lambda| \sin \phi}{1 +3 |\lambda|^2}.
\end{eqnarray}
The most accurate determinations of $|\lambda|$ (current world average
$|\lambda|= 1.2670 \pm 0.0035)$ come from measurements of
$A$ \cite{PDG99}.  The coefficients $a$, $A$, and $B$, respectively,
are measured to be $-0.102 \pm 0.005$, $-0.1162 \pm
0.0013$, and $0.983 \pm 0.004$ \cite{PDG99}.
Several previous experiments found the value of $D$, and thus
$\sin \phi$, to be consistent with zero at a level of precision well below
1\%.  The three most recent such
measurements found $D =( -1.1 \pm 1.7) \times 10^{-3}$ \cite{Stei74}
and $D = (2.2 \pm 3.0) \times 10^{-3}$ \cite{Eroz78}, and
$D = (-2.7 \pm 5.0) \times 10^{-3}$ \cite{Eroz74}, constraining $\phi$  to
$180.07^{\circ} \pm 0.18^{\circ}$ \cite{PDG99}.

Final state interactions give rise to phase shifts of the outgoing electron
and proton
Coulomb waves that are time reversal invariant but motion reversal
non-invariant.
Thus $D$ has terms that arise from phase shifts due to pure Coulomb
and weak magnetism
scattering.
The  Coulomb term vanishes in lowest order in V-A theory \cite{Jack57},
but scalar and tensor interactions could contribute. The Fierz interference
coefficient measurements \cite{Wenn68,Carn94} can be used in limiting this
possible contribution to
\begin{equation}
|D^{EM}|< (2.8 \times 10^{-5}) \frac{m_e}{p_e}.
\end{equation}
Interference between Coloumb scattering amplitudes 
and the weak magnetism amplitudes
produces a final state effect of order (${{E_e}^2}/{p_e m_n}$).  This weak
magnetism effect is predicted to be \cite{Call67}
  \begin{equation}
  D^{WM}=  1.1 \times 10^{-5}.
\end{equation}

The $D$ coefficient has also been measured for $^{19}$Ne decay, with
the most precise experiment finding $D_{Ne}= (4 \pm 8) \times 10^{-4}$
\cite{Hall84}.   The predicted final state effects for $^{19}$Ne are
approximately an order of magnitude larger than those for the neutron
and may be measured in the next generation of $^{19}$Ne experiments.  For
$^8$Li, a triple-correlation of nuclear spin, electron spin and
electron
momentum has  been measured, with the most precise measurement at
$R =( 0.9 \pm 2.2) \times 10^{-3}$ \cite{Srom96}.  Unlike $D$, a
nonzero $R$ requires the presence of scalar or tensor couplings and
thus is a tool to search for such couplings.  The  electric
dipole moments (EDMs) of the electron \cite{electronEDM},
neutron \cite{neutronEDM}, and
$^{199}$Hg atom \cite{HgEDM} are arguably the most
precisely-measured
$T$-violating
parameters and bear on many of the same theories as $D$.
Table \ref{Dlimits} summarizes the current constraints on $D$ from analyses
of data on other $T$-odd observables for the Standard Model
and
extensions \cite{Herc95}.  For lines 2-5 these limits are derived from the
measured neutron
or
$^{199}$Hg EDM.

\begin{table}[h]
\begin{center}
\begin{tabular}{ll}
Theory & $D$\\
\hline
1. Kobayashi-Maskawa Phase & $<10^{-12}$\\
2. Theta-QCD & $<10^{-14}$   \\
3. Supersymmetry & $\lesssim 10^{-7}-10^{-6}$ \\
4. Left-Right Symmetry & $\lesssim 10^{-5}-10^{-4}$ \\
5. Exotic Fermion  &  $\lesssim 10^{-5}-10^{-4}$ \\
6. Leptoquark & $\leq$present limit \\
\end{tabular}
\end{center}
\caption{Constraints on $D$ from analyses of other $T$-odd observables for
the Standard
Model and extensions.}
\label{Dlimits}
\end{table}
In the nearly two orders of magnitude between the present limit on
$D$  and the final state effects lies the opportunity to directly observe
or  limit new physics.  Moreover, accurate calculations of magnitude 
and energy dependence of the final
state   effects can be made to extend
the range of exploration still further.

\section{Overview of the emiT Detector}\label{overview}

\begin{figure*}
\begin{center}
\includegraphics[width=6.6in]{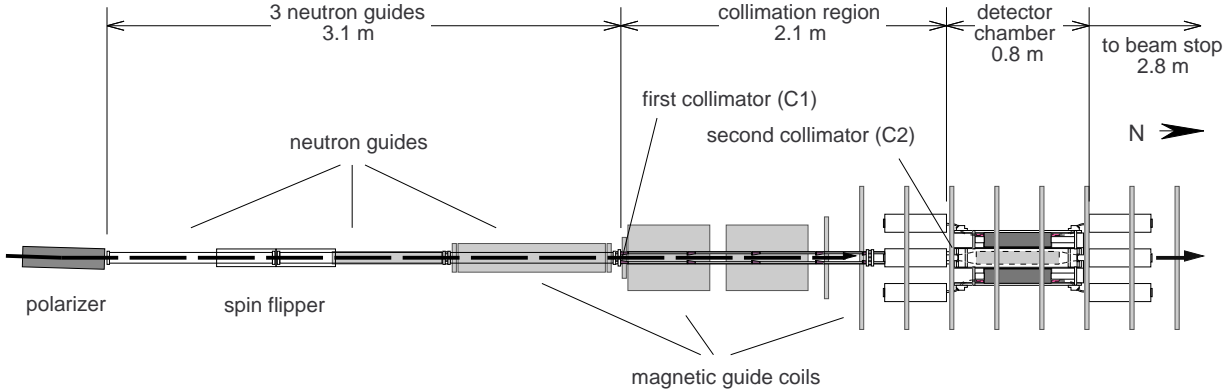}
\end{center}
\caption{The emiT experimental apparatus
beamline layout.  The neutrons traveled through 8.8 meters of guides
and vacuum components before reaching the beam stop.}
\label{beamline}
\end{figure*}

\begin{figure}[ht]
  \begin{center}
  \includegraphics[width=3.075in]{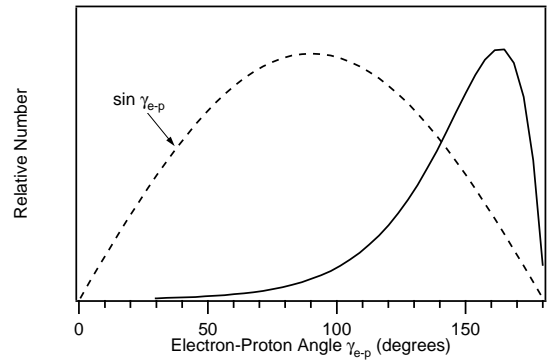}
\caption{
Although the cross-product (dashed line) is maximized at electron-proton
detection angles of 90$^\circ$, the overall sensitivity to $D$ (solid
line) is enhanced at larger angles due to the phase space for the
decay.  Placing the detectors at 135$^\circ$ allows for an octagonal
geometry that combines greater symmetry, acceptance, and sensitivity
when compared to placement of the detectors at 90$^\circ$.  The solid
curve in this figure is the sensitivity for a zero-radius beam, which would
exhibit a
factor of 7 enhancement for 135$^\circ$ as compared to 90$^\circ$.
For our nearly 3 cm-radius beam, the enhancement factor is close to 3.}
\label{coincang}
\end{center}
\end{figure}

In the emiT apparatus, a beam of cold neutrons is polarized and
collimated before it passes through a detection chamber with
  electron and proton detectors (four each).  A schematic of the experiment is
  shown in Figure \ref{beamline}. 
  The most
significant improvements over previous experiments are the
achievement of near-unity polarization ($>93\%$ compared to 70\% in
\cite{Stei74}) and the construction of a detector with greater acceptance and
greater sensitivity to the $D$ coefficient.  The octagonal arrangement
of the eight detector segments gives them nearly full coverage of the 2$\pi$ of
azimuthal angle around the beam, nearly twice the angular acceptance
in previous experiments, and the detector segments  are  longer
than in previous experiments.  The placement of the two types of detectors
at relative angles of 135$^{\circ}$ is also an improvement over
previous experiments, in which the coincidences were detected at
90$^{\circ}$.  While the cross product is greatest at 90$^{\circ}$,
the preference for larger
electron-proton angles in the decay makes placement of the detectors
at 135$^{\circ}$ the best choice to achieve greater symmetry, greater
acceptance, and greater sensitivity to $D$ (see Figure  \ref{coincang}).
 Combined with the higher neutron polarization from
the supermirror polarizer our geometry provides for an overall sensitivity
to $D$ that is a factor of $\approx 7$ greater than previous measurements,
assuming the same cold neutron beam flux.

  The first run of the experiment was conducted at
the NIST Center for Neutron Research (NCNR)
  in Gaithersburg, MD.
The experimental apparatus is outlined below, while more detailed
descriptions can be found in \cite{Lisi99,Hwan98}.

\subsection{Polarized Neutron Beam}\label{PolBeam}

The NCNR operates a 20-MW, heavy-water-moderated research reactor.
Neutrons from the reactor pass through a liquid hydrogen
moderator to make cold neutrons with an approximately
Maxwellian velocity distribution at a temperature of about 40 K.
The average neutron velocity is about 800 m/s.
The neutrons are transported 68 meters to the apparatus via
a  $^{58}$Ni-lined neutron guide.   Neutrons are totally internally 
reflected if they enter with an angle of incidence less than 2 mrad 
for each \AA~of de Broglie wavelength.
The capture flux of the neutrons  was measured using a gold
foil activation technique to be $\rho_n v_0 = 1.4\times 10^9$n/cm$^2$/s
 (where $v_0$ = 2200 m/s) at the end
of the neutron guide.  (The capture flux quantifies the neutron density
 in the detector for the polychromatic beam.)
The beam passes through a cryogenic
beam filter of 10-15 cm of single crystal bismuth which  filters out residual
fast neutrons  and gamma-rays.

The neutrons are polarized with a double-sided
bender-type supermirror polarizer obtained from the
Institut Laue-Langevin in Grenoble, France
  \cite{Scha89}.
   The supermirror  consists of 40
Pyrex \cite{BrandDisclaimer} plates coated on both sides with cobalt, titanium,
and gadolinium layers that maximize
the reflection of neutrons with the desired spin state while
absorbing nearly all neutrons of the opposite spin state.
The supermirror was measured to polarize a 4.5 cm by 5.5 cm beam with 24\%
transmission relative to the incident unpolarized flux. The neutron
polarization
was determined to be $>93\%$ ($95\%$ CL).

\begin{figure}
  \begin{center}
  \includegraphics[width=3.375in]{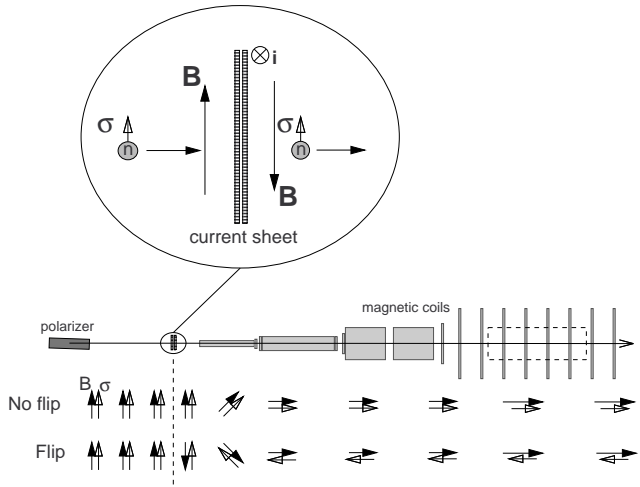}
  \caption{Two  sheets of current-carrying
  wires create a magnetic field of opposite orientation on each
  side.  The field orientation changes so rapidly that the spin of a neutron
  passing through the current sheets cannot follow the field reversal, and
  the neutron polarization is reversed with respect to the magnetic guide
field.
Downstream the magnetic field and polarization are rotated
  adiabatically from transverse in orientation to longitudinal.
  }
  \label{spintrans}
  \end{center}
  \end{figure}

The neutrons travel the one meter from the polarizer to the spin-flipper
inside a Be-coated glass flight tube in which a small helium
overpressure is maintained to minimize beam attenuation via air
scattering.
The neutrons, which have spins that are transverse to their motion, then pass
through two layers of aluminum wires which comprise the current-sheet spin
flipper. When the current in the second layer is antiparallel to that in the 
first there is no net magnetic field and the neutron polarization is
unaffected.  When the currents are parallel, the neutron spin does not
adiabatically follow
the rapid change in field orientation and thus the sense of $\langle \mathbf{J}\rangle \cdot
\mathbf{B}$ is
reversed.
 Downstream of the spin flipper, weak magnetic fields adiabatically rotate
  the spin to longitudinal, {\it i.e.} parallel or antiparallel to the
  neutron momentum.
  The longitudinal guide fields are 2.5 mT upstream and 0.5 mT within the
detector.
Figure \ref{spintrans} shows the spin transport system.
The polarization direction is reversed every 5
seconds.
In the detection region, the longitudinal field is produced by eight
50 amp-turn current loops of 1 m diameter.  The loops are aligned
 to within 10 mrad of the detector axis using a sensitive
  field probe and an AC lock technique.  Additional coils canceled
 the transverse components of the Earth's field and local gradients of 7.5
$\mu$T/m.

The vacuum chamber begins at the spin flipper with two meters of 
Be-coated flight tubes, through which the neutrons travel toward the
collimator region.
Two collimators of 6 cm and 5 cm diameter openings separated by 2 m
define the beam.  These and  5 additional ``scrapers'' between them
  consist of rings of $^6$LiF which absorb the neutrons.  Behind each
  ring is a thick ring of high-purity lead
which absorbs the gamma-rays from the reactor and those produced by
  neutron captures upstream.   Between scrapers,
the walls of the beam tube are lined with $^6$Li-loaded glass to
  absorb stray neutrons.

\begin{figure}
  \begin{center}
  \includegraphics[width=3.075in]{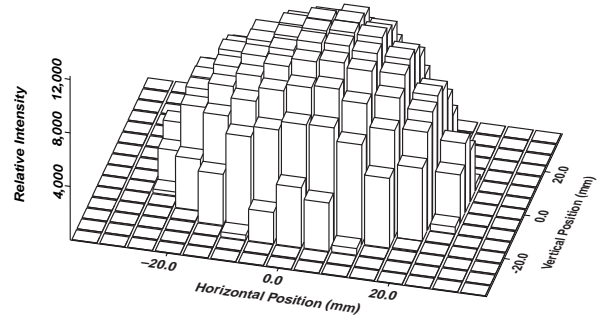}
\vspace{1 mm}
\caption{Neutron beam intensity profile at the entrance to the 
detection region obtained
from a scan across the beam face.}
\label{beamintens} \end{center}
  \end{figure}

A fission chamber mounted behind a sheet of $^6$Li-glass with a
1 mm pinhole aperture was scanned across the beam to obtain a
cross-sectional profile of the intensity as shown in Figure
\ref{beamintens}.  The neutron intensity was measured before and after
the experiment.  To determine the polarization at the entrance to the
detector, the beam passed through a second, single-sided, analyzing
supermirror
directly in
front of the scanning detector, and the ratio of intensities with the spin
flipped and
unflipped was measured.  The resulting flippng ratio measures a combination
of the
neutron-spin-dependent transmission efficiencies of the two supermirrors
and the neutron
spin flipping eficiency. From this, and assumptions about the spin flipping
efficiency,
we can determine the product of polarization efficiencies for the two
supermirrors
(polarizer and analyzer).  When the upper limit of 100\% spin flipping efficiency is
used, a lower
limit of the neutron beam polarization of $93\%$ ($95\%$ CL) is 
found.  This lower limit also includes the assumption that the 
flipping ratio for a pair of supermirrors identical to our analyzer would be
less than that of a pair of supermirrors identical to our polarizer by a
factor of
$2\pm 0.5\%$\cite{Scha89}.

   Downstream of the detection region the vacuum chamber diameter
increases to 40.6 cm, terminating with a $^6$Li-glass beam stop
2.8 m from the end of the detector.
  A 1 mm diameter pinhole at the
center of the beamstop allows about 1\% of the beam to pass
  through a silicon window into  a
  fission chamber detector that continuously monitors the neutron flux.

\subsection{Detector System}

\begin{figure*}
\begin{center}
\includegraphics[width=5.6in]{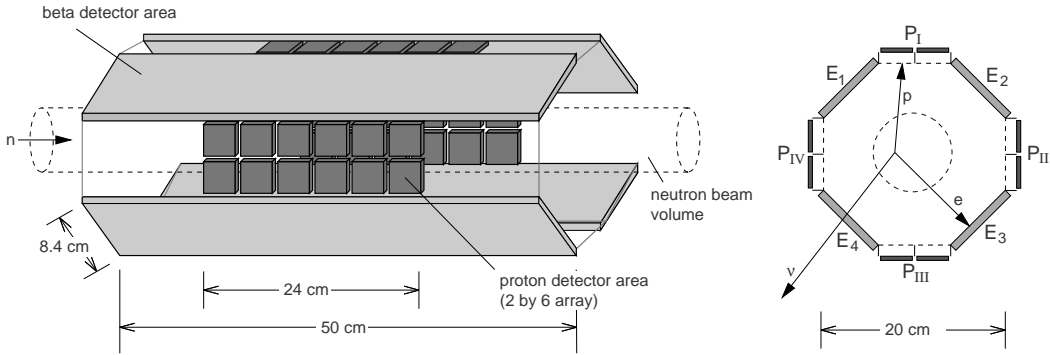}
\caption{Basic detector geometry
- an octagonal array of four each proton and electron detectors.}
\label{detschem}
\end{center}
\end{figure*}

Eight detectors surround the beam, each 10 cm from the beam axis as
shown in Figure \ref{detschem}.
The octagonal geometry places  electron and proton detectors at relative
angles of 45 and 135 degrees.  Coincidences are counted between detectors
at relative angles of 135 degrees.

\subsubsection{Electron Detectors}

The electron detectors are slabs (8.4 cm x 50 cm x 0.64 cm) of BC408
plastic scintillator  connected on each end to curved lucite
light-guides that channel the light to Burle 8850 photomultiplier tubes.
Each photomultiplier tube is surrounded by a mu-metal magnetic shield and
 a pair of nested solenoids acting as an active magnetic shield.
This combination of active and passive magnetic shielding
had a factor of 10 less impact (0.5
   $\mu$T) on the guide field at the beam center than the mu-metal alone.

\begin{figure}
\begin{center}
\includegraphics[width=3.225in]{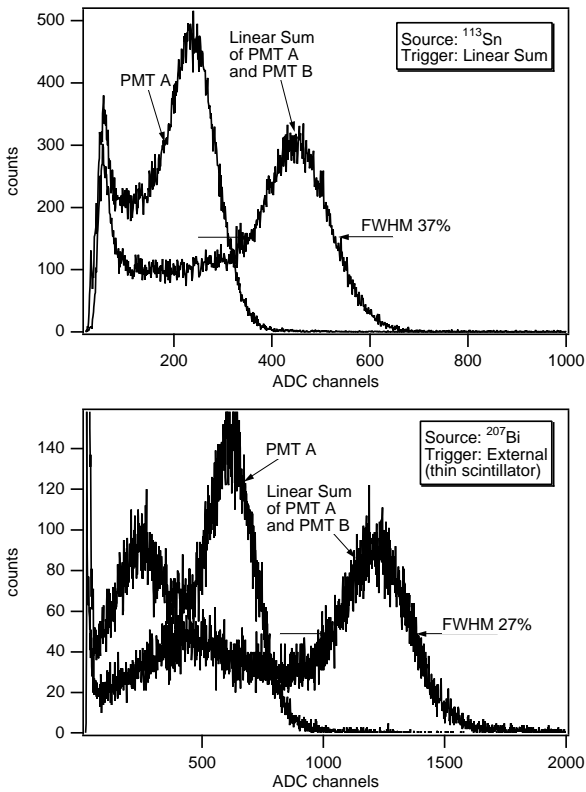}
\caption{Spectra produced during the energy calibration of the electron
detectors.  Shown are histograms of the charge collected in an individual
phototube (PMT A)  and of the total charge in the analog hardware
sum of the two phototube signal.  The spectrum in PMT B (not shown) is
roughly identical to that in PMT A. For the $^{113}$Sn, a level-crossing
discriminator triggers on the analog sum signal.  The peak visible is
the 364 keV conversion electron.  To suppress the contribution of
gamma-rays in the   $^{207}$Bi spectra the data acquisition is
triggered on   a thin scintillator placed between the source and the detector.
   The largest feature in these spectra falls at an energy of 882 keV
(the 976 keV conversion electron energy minus the energy loss in the thin
scintillator.) }
\label{betaspecs}
\end{center}
\end{figure}

The scintillator thickness of 0.64 cm is just greater than that necessary
to stop the
most energetic (782 keV) of the electrons from neutron.  The
scintillators are wrapped with aluminized mylar and aluminum foil to
prevent charging and to shield the detectors from x-rays and
field-emission electrons in the vacuum chamber.   For each segment, the
energy response was calibrated with cosmic-ray muons and
conversion electrons from $^{207}$Bi and $^{113}$Sn (see Figure
\ref{betaspecs}.)

\subsubsection{Proton Detectors}

\begin{figure}
\begin{center}
\includegraphics[width=2.375in]{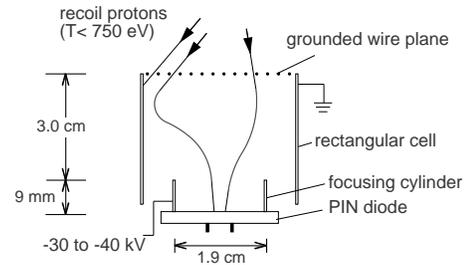}
\caption{{Geometry of the electrodes that
accelerate and focus the  protons onto a PIN diode.}}
\label{focus}
\end{center}
\end{figure}

  Each proton detector has  an array of 12 PIN diodes
of 500 micrometer thickness arranged in two rows of 6.
  The diodes are held within a stainless steel high voltage
electrode.  Over each diode an open cylinder protrudes from the face
of the electrode, shaping the field to focus and accelerate the
protons as shown in Figure \ref{focus}.
Thus each diode collects protons focused from a region 4 cm  $\times$ 4
cm even though it has an
 active area of only 1.8 cm $\times$ 1.8 cm.
   The diodes and their electronics are held at -30 to -40 kV.
Between the electrode and the beam is a frame strung with
  80  0.08-mm
gold-plated tungsten wires that define a plane of
electrical ground.
   Protons drift in a field-free region until they
pass this plane, and then are accelerated by the high voltage and
focused onto the nearest PIN diode below.
Near both ends of the detector array are two cryopanels held at
liquid nitrogen temperature.  Water vapor,  released
predominantly by the
scintillators and other plastic components, is pumped onto the
cryopanels to prevent condensation on the cooled PIN diodes.

\begin{figure}
\begin{center}
\includegraphics[width=3.375in]{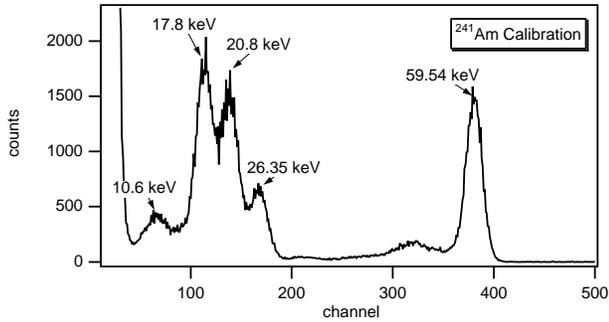}
\caption{Energy calibration spectrum of a PIN diode detector using an
$^{241}$Am source. The FWHM is 2.9 keV at the 59.5~keV line.}
\label{pcalib}
\end{center}
\end{figure}

The charge in the PIN diode produced by each proton is amplified by 10
V/pC with a preamplifier mounted directly behind the PIN diode.  These
circuits and the PIN diodes are cooled with liquid nitrogen to about
0$^{\circ}$C to decrease
electronic noise.  Preamplifier signals are processed in a custom
VME-format shaper/ADC board with programmable gain and operating mode
parameters.  The PIN diodes were calibrated with x-rays from an
$^{241}$Am source as shown in Figure \ref{pcalib}.

\subsubsection{Background}

The  background in the detectors was primarily related to the beam or
to the high voltage bias.  Closing the beam shutter upstream of the neutron
filter stops virtually all neutrons and about 1/3 of the gamma-rays
coming from the reactor along the beamline.
With the  shutter closed, the rates in each  detector were less than 100
Hz, primarily from dark current, reactor gamma-rays, and cosmic rays.
With the
shutter open, the detectors see an increased gamma-ray flux primarily from
neutron
captures in the apparatus, triggering the detectors at less than 1 kHz per
electron
detector and less than 1 kHz for all PIN diodes combined.   This
results in deadtime  less than  3\% for the beam-related background.
  At its worst, the high-voltage-related background, consisting of x-rays,
light,
electrons, and ions, led to rates in the hundreds of kHz in the
detectors.  It was
  reduced at times by conditioning and cleaning of electrodes but
  varied by orders of magnitude during the run.

\subsection{Data Acquisition}

\begin{figure}
\includegraphics[width=3.375in]{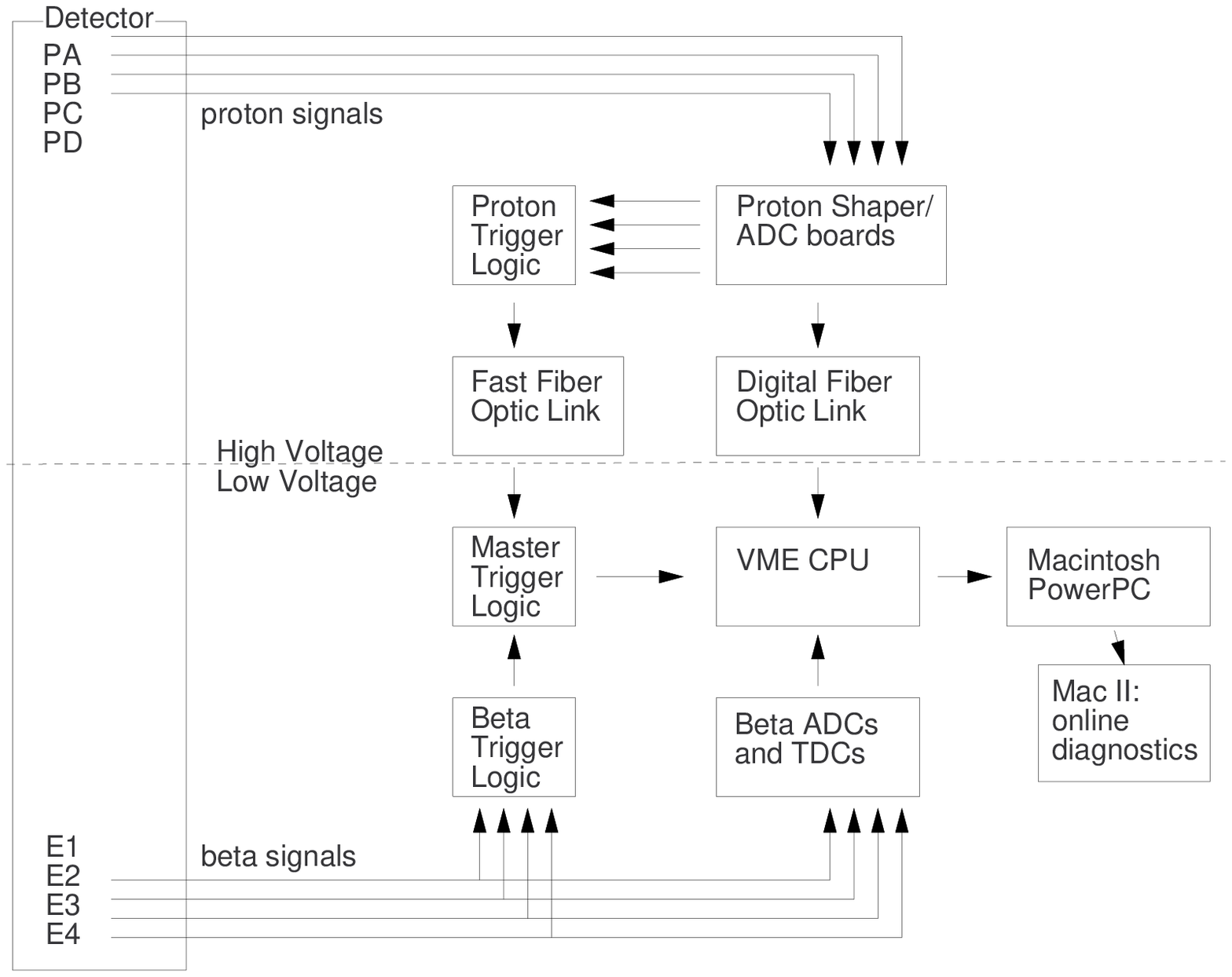}
\caption{Data acquisition components.}
\label{DAQ}
\end{figure}

A block diagram of the data acquisition system is shown in Figure
\ref{DAQ}.
The identification of neutron decay events is simplified by the fact that
the proton signal is observed 0.5 $\mu$s to 2 $\mu$s after the electron signal.
 The recoil protons, with maximum energies of
only 750 eV, require this time to drift from the point of decay to the
face of the proton detector.
Events are accepted by the coincidence trigger when the electron signal
arrives within a coincidence time window $\pm \tau_{coinc}/2$ of a proton
signal.   The duration
  $\tau_{coinc}$ of this window was originally 14 $\mu$s and was shortened to 7
$\mu$s midway through the experiment to reduce the system deadtime.
Each stored event contains the location (PIN diode) and energy for the proton
event, location (electron detector), energy of the electron event,
relative time
between
individual  signals  from the two phototubes in the electron detector,
  relative time of arrival of the proton and electron
signals, and the orientation of the neutron polarization.
Every 30 seconds during the data collection, information is recorded
from the system monitors which include system livetime,
magnet currents, neutron flux at the beam stop, vacuum pressure,
proton detector high voltage, and high voltage leakage current.
Periodically, the data acquisition collects singles spectra from
all of the individual detectors.

\section{Experimental Run}

\subsection{Data Collection}

The experiment was installed at the NCNR during December 1996 and
January  1997.  From February through August 1997, 50 GB of data were
collected and stored. The data are divided into 626 files representing
continuous runs, typically four hours in duration.  These are grouped
into 125 series, within which running conditions varied little.
For one week in August
  a systematic test was run in which the beam was
  distorted and the polarization guide field direction changed.  The
  purpose and results of this test will be described in Section
  \ref{Anal}.

\begin{figure}
\includegraphics[width=3.375in]{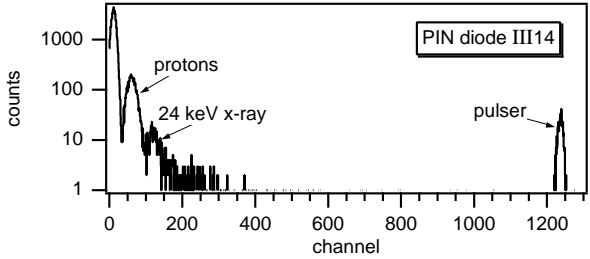}
\caption{Energy spectrum in PIN diode III14, near which is mounted a
weak $^{119}$Sn source producing a 24 keV x-ray.
The protons, accelerated to 36 keV but
measured at less than 20 keV, are visible between the background and
the x-ray peak.  The peak on the far right from a low-rate pulser input
directly into the preamplifier is used to monitor gain and resolution.}
\label{C14spect}
\end{figure}

Instabilities in the proton detector high voltage  made it impossible
to operate all channels of the detectors at all times.
Sometimes the
electrodes simply would not hold the necessary voltage, and at other
times a large spark or series of sparks would damage the electronics
held at high voltage.  Less than half of the data
were collected when all four proton detectors were functioning.
Another limitation to the detector uniformity were variations in the
measured proton energy deposited in the PIN diodes.  In preliminary tests,
  the  surface dead-layers of the PINs were
measured to be 20$\pm 2$ $\mu$g/cm$^2$ as specified by the
manufacturer, Hamamatsu.
In a dead-layer of this thickness a 35 keV
proton loses 10 keV of energy.  The proton energies measured during
the experiment, however, were 12-18 keV, an average of 20 keV below the energy
imparted to them by acceleration through 34-38 kV (see Figure
\ref{C14spect}.)  With widths
(FWHM) of approximately 10 keV,
  these peaks are not well separated from the background.
High background rates necessitated the setting of thresholds at levels 
such that some neutron decay events were also rejected.  This and the 
data acquisition deadtime were the primary limitations to the 
statistics of the experiment.  
A deadtime per event of 2 ms was
necessary for stability of the system.
Even with the reduction in length of the coincidence
window, the high rate of background kept the system at 40-60\%
deadtime for most of the data collection period.

\subsection{Event Selection}

\begin{figure}
\includegraphics[width=3.375in]{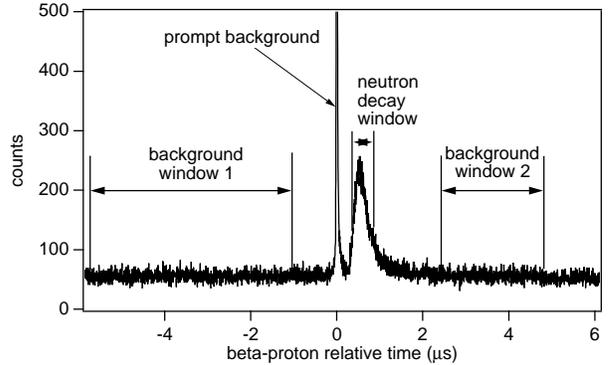}
\caption{{Time windows used to find the signal and estimate the
background. }}
\label{newTspect}
\end{figure}

Figure \ref{newTspect} shows an example of the relative time spectrum
for the coincidence data.  The large center spike, originating mainly
from multiple gamma rays produced by neutron captures in the
apparatus, defines zero time difference.  The neutron decay events are
accepted within a window 0.35 $\mu$s  to 0.9 $\mu$s after the prompt peak.
This window contains the majority of the neutron decay protons, while
excluding the tail of the prompt peak and the low-signal-to-background
tail of the proton peak. The background
to be subtracted from these events is estimated using the
rates in regions to either side of the decay and zero-time peaks.
Events are also selected on the basis of measured proton energy to
reduce the amount of background to be subtracted. The energy range
accepted is chosen solely by minimizing the fractional statistical uncertainty
  in the number of neutron decay events for each PIN diode-electron
detector pair.
  Specifically, if $N_{\Delta}$ is the number of coincidences counted
  by subtracting the background from the coincidences in the  0.35 to
  0.9 $\mu$s window, the
  energy range is chosen to minimize
\begin{equation}
\frac{\sigma_{N_{\Delta}}}{N_{\Delta}} \approx
   \frac{\sqrt{1+1/f}}{\sqrt{N_{\Delta}}},
   \end{equation}
   where $f$ is the signal to background ratio in this energy range.
  This increases the overall signal
to background on the 15 million good events from 0.8 to 2.5.

\section{Data Analysis and Uncertainty Estimation}\label{Anal}

\subsection{Determination of $D$ from Coincidence Events}

For each PIN diode-electron detector pair in a given data series, the count
rate can be expressed as
\begin{eqnarray}
N_\pm^{\alpha i} = N_0\epsilon^\alpha\epsilon^i
\big [ 
    K^{\alpha i}_1+a K^{\alpha i}_a \pm P{\hat \sigma} \cdot
                 \nonumber\\
  (A {\mathbf K}^{\alpha i}_A + B {\mathbf K}^{\alpha i}_B+ D {\mathbf
K}^{\alpha i}_D ) \big ],
\label{eq:counts}
\end{eqnarray}
where $N_0$ is a constant proportional to the beam flux,
the $\epsilon^\alpha$ and $\epsilon^i$ are detector efficiencies for a PIN
diode and electron detector
respectively. The average of the neutron polarization vector
over the detector volume, given by $P\hat\sigma$,is assumed to be uniform and constant over 
time, lying along the direction of the 0.5 mT guide field.
The $\pm$ signs correspond to the two signs of the polarization.
The factors $K_1^{\alpha i}$ and $K_a^{\alpha i}$ are geometric factors
derived from Equation \ref{dW(ps)} by integrating $1$ and
${{\mathbf p}_e \cdot {\mathbf p}_\nu}/E_e E_{\nu}$, respectively, over the
beta-decay phase space, the neutron beam volume, and the acceptance
of each electron-detector--PIN-diode detector pair.
Similarly, the factors ${\mathbf K}_A^{\alpha i}$, ${\mathbf K}_B^{\alpha i}$,
 and ${\mathbf K}_D^{\alpha i}$,
are obtained by integrating the vectors: ${{\mathbf p}_e}/{E_e}$,
${\mathbf p}_\nu /{E_\nu}$, and
$({\mathbf p}_e \times {\mathbf p}_\nu)/ {E_e E_{\nu}}$.

We produce the following efficiency-independent asymmetries
\begin{equation}
w^{\alpha i} =
{N_+^{\alpha i} - N_-^{\alpha i}  \over
   N_+^{\alpha i} + N_-^{\alpha i}    }.
\label{eq:asymm}
\end{equation}
   From Equation~\ref{eq:counts} we get
\begin{equation}
w^{\alpha i}=
 P {\hat \sigma} \cdot (A {\tilde {\mathbf K}}^{\alpha i}_A + B {\tilde
{\mathbf K}}^{\alpha i}_B +
                       D {\tilde {\mathbf K}}^{\alpha i}_D ),
\end{equation}
where we use the definitions
\begin{equation}
{\tilde {\mathbf K}}^{\alpha i}_A = \frac{{\mathbf K}^{\alpha i}_A}
{K^{\alpha i}_1 + a K^{\alpha i}_a}\ \ {\rm etc.}
\end{equation}

\begin{figure}
 \begin{center}
\includegraphics[width=1.3 in]{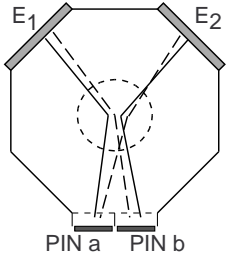}
\caption{{The data from two PINs at
the same $z$-position in a proton segment can be used to cancel the
effects due to the electron and neutrino asymmetries. The coincidences
shown by solid lines (${\rm E}_1 {\rm PIN}_a$ and
${\rm E}_2 {\rm PIN}_b$) have approximately the same angle,
a little less than 135$^{\circ}$.
These are referred to as ``small-angle'' coincidences.   The
``large-angle'' coincidences for this pair of PINs
(${\rm E}_1 {\rm PIN}_b$ and ${\rm E}_2 {\rm PIN}_a$) are the dashed lines.}}
\label{TwoPIN}
\end{center}
\end{figure}

Consider the two detector pairings PIN$_a$-E$_1$ and PIN$_b$-E$_2$
indicated in Figure~\ref{TwoPIN}. The corresponding values of
${\mathbf K}_D^{\alpha i}$ have opposite sign while ${\mathbf K}_A^{\alpha
i}$ and ${\mathbf K}_B^{\alpha i}$  have the same sign. We therefore
combine asymmetries from two proton-electron detector pairings to produce
the combination
\begin{eqnarray}
v^{b2:a1} = \frac{1}{2} [ w^{b2} - w^{a1} ]\\
           = \frac{1}{2} P \hat{\sigma} \cdot \Big [ D \big ({\tilde{\mathbf
K}}^{b 2}_D
-{\tilde{\mathbf  K}}^{a  1}_D \big )\nonumber \\
+ A \big ( {\tilde{\mathbf  K}}^{b  2}_A - {\tilde{\mathbf K}}^{a  1}_A
\big ) +
    B \big ({\tilde{\mathbf  K}}^{b  2}_B - {\tilde{\mathbf K}}^{a  1}_B \big)
    \Big ].
\label{eq:kpd}
\end{eqnarray}
For uniform detection efficiency the difference $({\tilde{\mathbf
K}}^{b 2}_D - {\tilde{\mathbf  K}}^{a
1}_D \big )$
lies along the detector axis, $\hat z$, while the differences
$({\tilde{\mathbf  K}}^{b 2}_A - {\tilde{\mathbf  K}}^{a 1}_A \big )$  and
$({\tilde{\mathbf  K}}^{b 2}_B - {\tilde{\mathbf  K}}^{a 1}_B \big )$
lie perpendicular to the detector axis. For
a polarized neutron beam with perfect cylindrical symmetry aligned with the
detector axis,
$\tilde{\mathbf  K}^{b2}_D\cdot \hat{z} =
 - \tilde{\mathbf  K}^{a1}_D\cdot \hat{z}$, and
\begin{equation}
v^{b2:a1} =  P D{\tilde{\mathbf  K}}^{b2}_D \cdot \hat{z} = - P
D{\tilde{\mathbf  K}}^{a1}_D \cdot \hat{z}.
\label{eq:kpd2}
\end{equation}
Departures from perfect symmetry and perfect alignment of the neutron
polarization require that the $A$ and $B$
correlation terms be retained in Equation \ref{eq:kpd}. The resulting
systematic effects are
discussed in Section \ref{sec:FalseD}.

Additionally, as shown in Figure \ref{TwoPIN}, there are two classes
of electron-PIN pairs: those that make an angle smaller than 135$^{\circ}$
($b2:a1$) or an angle larger  than 135$^{\circ}$ ($a2:b1$).
We thus separate our data into a
{\em small-angle} group and a {\em large-angle} group
giving two statistically independent results for each
PIN-diode--electron-detector pairing.

\subsection{Monte Carlo Methods}

We use two Monte Carlo calculations to determine the values of
${K_1}$, ${K_a}$, ${{\mathbf  K}_A}$, ${{\mathbf  K}_B}$,
and ${{\mathbf  K}_D}$.
The results from these two completely independent calculations
are in excellent agreement.
In both calculations neutron decay events are generated randomly within a
trapezoid-cylindrical geometry ({\em i.e.} a tube with
divergence) that can be offset with respect to the
detector axis.
A realistic beam profile, representative of Figure \ref{beamintens}, can be modeled by combining results from several
different
trapezoids.
In one of the Monte Carlo calculations the tracking of protons
and electrons is done
with the CERN Library GEANT3 Monte Carlo package \cite{GEANT},
while in the other tracking is implemented within
the code itself. In both, the emiT detector geometry is specified
with uniform efficiency over the active area of each scintillator and 
over
the square focusing region of each PIN diode.

The constants defined in Equation \ref{eq:counts} are given by
\begin{equation}
K^{\alpha i}_x = \sum \delta^{\alpha i} X
\end{equation}
where $X=1$,
${{\mathbf p}_e \cdot {\mathbf p}_\nu}/E_e E_{\nu}$,
${{\mathbf p}_e}/{E_e}$,
${{\mathbf p}_\nu}/{E_\nu}$, and
${{\mathbf p}_e \times {\mathbf p}_\nu}/ {E_e E_{\nu}}$
for $x=1,~a,~A,~B,$ and $D$, respectively.
We have studied systematic uncertainties associated with
potential non-uniformities in the beta efficiencies and
included them in the final uncertainty for the constants $K^{\alpha i}_x$.
These constants (a total of 11, taking into account the three
directions for each vector) are accumulated in a file
that is read to calculate the factors $v$ (Equation \ref{eq:kpd}) for
different orientations of the polarization.

Values of $|{\tilde{\mathbf K}}^{\alpha i}_D \cdot {\hat z|}$
are used directly in the interpretation of the result
for $D$.  Variations among the PIN diode pairs of individual values of
${\tilde{\mathbf K}}^{\alpha i}_D$
within a given proton segment are negligible, and average values
($|\tilde{\mathbf K}_D\cdot \hat{z}|$) can be used.
They are found to be  $ 0.424 \pm 0.010$  and
$0.335 \pm 0.020$,
for the small- and large-angle coincidences, respectively.
The uncertainties are primarily from uncertainties in the
geometry of the beam.  Values for the other $K^{\alpha i}_x$ are used in
the estimation of systematic uncertainties described  in the following section.

\subsection{Discussion of Systematic Uncertainties}
\label{sec:FalseD}

The largest of the systematic effects can be shown to be the contributions to
the $v$ (Equation \ref{Dtransverse}) that arise due to
the misalignment of the neutron polarization with respect to the detector axis.
A transverse component of the
polarization produces a significant contribution to $v^{b2:a1}$ because  the
vector differences ${{\tilde{\mathbf  K}}^{b 2}_A}-{\tilde{\mathbf  K}}^{a
1}_A$
and ${{\tilde{\mathbf  K}}^{b 2}_B}-{\tilde{\mathbf  K}}^{a 1}_B$
are predominantly perpendicular to the detector axis.
(For example, ${{\tilde{\mathbf  K}}^{b 2}_A}-{\tilde{\mathbf  K}}^{a 1}_A$
is proportional to
the integral of ${\mathbf{p}}_e(E_1)-{\mathbf{p}}_e(E_2)$ and is directed
horizontally
 to the left in Figure \ref{TwoPIN}.  The difference
${{\tilde{\mathbf  K}}^{b 2}_B}-{\tilde{\mathbf  K}}^{a 1}_B$ is
antiparallel to
${{\tilde{\mathbf  K}}^{b 2}_A}-{\tilde{\mathbf  K}}^{a 1}_A$.)
For an azimuthally symmetric neutron beam, it can be shown that
for each proton detector segment (labeled with subscripts $\eta$= I, II, III, IV) the
weighted average of the $v^{\alpha i:\beta j}$ for all large or small
detector pairs can be
expressed as
\begin{equation}\label{Dtransverse}
v_\eta^{l/s} = PD(\tilde{\mathbf K}_D^{l/s}\cdot\hat\sigma) +
\alpha_\eta^{l/s} \sin
\theta_{\sigma} \sin
(\phi_{\eta}-\phi_{\sigma}),
\end{equation}
where $\theta_{\sigma}$  and $\phi_{\sigma}$  are the polar and azimuthal
angles of
$\hat\sigma$, and $\phi_\eta$=0$^\circ$, 90$^\circ$, 180$^\circ$, and
270$^\circ$
respectively for detectors I, II, III, and IV.
This dependence can be
derived analytically for zero beam radius and is confirmed
by Monte Carlo simulations
 for  symmetric beams of finite radius.
The coefficients $\alpha_\eta$ measure
the combined effects
of the $A$ and $B$ correlations for each proton detector segment.

If the symmetry of the four
sets of proton detectors were perfect, {\it i.e.} $\alpha_{\rm
I}=\alpha_{\rm II}=\alpha_{\rm III}=\alpha_{\rm IV}$,  the contributions
due to the
$A$ and $B$ coefficients would average to zero, and Equation \ref{eq:kpd2}
would be valid, even with a polarization misalignment.
In the absence of perfect symmetry, these contributions do not cancel when
the four proton detectors are combined, and a false $D$ contribution would
result from
the application of Equation \ref{eq:kpd2}. This false
$D$ is proportional to the product of two effects that are both small: the
misalignment of the neutron polarization with respect to the detector axis
($\theta_\sigma$)
and the departure from perfect symmetry of the proton detectors
($\Delta\alpha = 1/2(\alpha_{\rm I}-\alpha_{\rm III})
+1/2(\alpha_{\rm II}-\alpha_{\rm IV})$).
Such an effect is
called the ``tilting asymmetric transverse polarization'' effect, or
``Tilt ATP''\cite{Eroz74,Wass94}.

\begin{figure}
\includegraphics[width=3.375in]{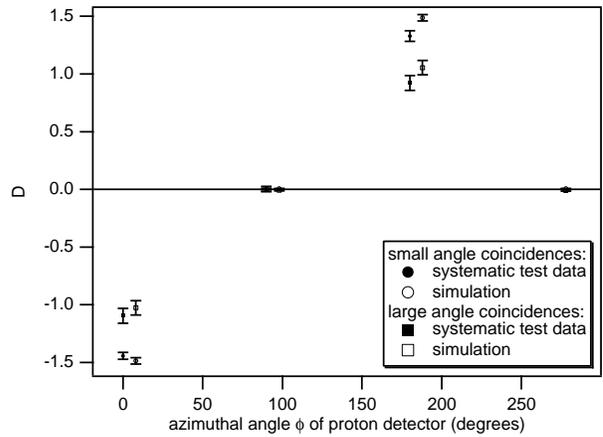}
\caption{During the  systematic test
  with a  90$^{\circ}$  polarization tilt,  the
false $D$ in each proton segment is clearly visible.
  (The proton detector segment IV  at $\phi_{\rm IV}$ = 270$^{\circ}$ was not
operational.)
Also shown  are the
results of Monte Carlo simulations of the systematic test
conditions.  The error bars shown are purely statistical, and
are not an accurate estimate of the total uncertainty in the calculation,
of which the
largest contribution is uncertainty about the shape of the beam.}
\label{fg:ATPDs}
\end{figure}

The ATP effect was intentionally amplified for a systematic test, run
with transverse polarization ($\theta_{\sigma}=90^{\circ}$
$\phi_{\sigma}=\phi_{\rm IV}= 270^{\circ}$) and a distorted
neutron beam. The neutron beam was
distorted by blocking half of the beam with a neutron absorber placed
upstream near the spin flipper.
The results of this test are shown in Figure \ref{fg:ATPDs}.
This demonstration that the experiment can measure an asymmetry consistent with
the Monte Carlo calculation serves as a strong check on both the operation
of the detector and the validity of the analysis method.

A false $D$ also arises if the  polarization has transverse components not
described by a simple
tilt. The form of Equation \ref{Dtransverse} shows that a net azimuthal
component of $\hat\sigma$ also
results in a contribution to $v_\eta$ that does not average to zero when
data from
proton segments I-IV are combined.
This effect, referred to as a ``twisting
asymmetric transverse
polarization'' (``Twist ATP'') is shown by Monte Carlo simulations to be
less than
$10^{-4}$ for azimuthal polarizations of less than 1 mrad.  For this
reason, all sources of guide field distortion are kept to less than
1 mrad, and materials of low magnetic permeability
(less than 0.005 $\mu_0$) were used in the detection region.  There are
 exceptions to this requirement, however the net effect of all
additional permeability was measured to produce less than 1 mrad of
distortion of the guide field anywhere in the detector region.

Variations in the neutron flux ($\Phi$) and polarization ($P$) that depend
on neutron helicity
 yield a false $D$. For this experiment the effects due
to misalignment of the neutron spin are small\cite{Dflasefootnote},
so that these systematic effects, to first order in
$\Delta \Phi/\bar{\Phi}$ and $\Delta P$ are
\begin{equation}
D_{false}(\Delta \Phi) = \frac{\Delta \Phi}{\bar{\Phi}} PD
{\hat \sigma} \cdot (A\langle{{\tilde{\mathbf  K}}_A}\rangle+B\langle{{\tilde{\mathbf  K}}_B}\rangle),
\end{equation}
and
\begin{equation}
D_{false}(\Delta P) =
\Delta PD{\hat \sigma} \cdot (A\langle{{\tilde{\mathbf
K}}_A}\rangle+B\langle{{\tilde{\mathbf  K}}_B}\rangle).
\end{equation}
Here $\Delta \Phi = \Phi_{\uparrow}-\Phi_{\downarrow}$,
and  $\Delta
P = P_{+} -
P_{-}$. ($P$ in Equations \ref{eq:kpd} and
\ref{eq:kpd2} would be replaced by $\bar{P}=(1/2)[P_{+} + P_{-}]$.)  
The
$\langle{{\tilde{\mathbf  K}}_A}\rangle$ and $\langle{{\tilde{\mathbf  K}}_B}\rangle$ are average
values for all
PIN-diode--electron-detector pairings.

Our data
provide an upper limit of 0.002 for $P{\hat \sigma} \cdot (A\langle{{\tilde{\mathbf
K}}_A}\rangle+B\langle{{\tilde{\mathbf  K}}_B}\rangle)$. We combine this with
neutron flux monitor data for $\Delta\Phi/ \Phi<0.004$, concluding
that $D_{false}(\Delta \Phi) < 8\times 10^{-6}D$.
The flipping ratio measurement has been used to derive a lower limit on
the spin flipper efficiency
of 82\% so that $\Delta P < 0.2$, and $D_{false}(\Delta P)  < 4\times
10^{-4} D$. We conclude that both effects are negligible in this measurement.

\subsection{Results}

\begin{figure}
\includegraphics[width=3.375in]{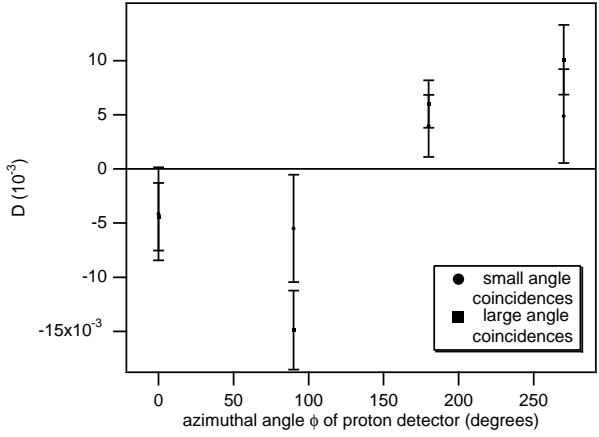}
\caption{{Results from each of the four proton segments for small-angle and
large-angle PIN-electron detector pairs. $D_\eta^{l/s}= v_\eta/(P\tilde{\mathbf
K}_D^{l/s}\cdot \hat z )$.
  Error bars are statistical.}}
\label{8Ds}
\end{figure}

A final value of $D_\eta^{l/s}= v_\eta/(P\tilde{\mathbf K}_D^{l/s}\cdot
\hat z)$
is found separately for large angle and small angle pairings
of each proton segment. The quantities $v_\eta$ are the weighted averages
of all PIN-electron detector pairs, $v^{l/s}(\alpha i:\beta j)$, within
each proton detector segment. Use of the weighted averages is justified
because
the systematic uncertainties described above have negligible variations
among the PIN diode pairs in a given detector.
The individual proton segment data ($v_\eta$) are then combined in an
arithmetic average so that
the sinusoidal variation given in Equation \ref{eq:kpd2} cancels to first
order in misalignments, {\it i.e.}
\begin{eqnarray}
 \sum_{\eta={\rm I}}^{\rm IV} v_\eta^{l/s} = 4D^{l/s} (P{\tilde{\mathbf
K}_D^{l/s}\cdot \hat z }) + {\cal O}(\theta_\sigma\Delta \alpha).
\label{eqcomb}
\end{eqnarray}
The error for $D^{l/s}$ includes the uncertainty in
the values of  ${{\tilde{\mathbf  K}}^{l/s}_D}\cdot \hat{z}$.
\begin{eqnarray}
\sigma_{D^{l/s}}^2
=\frac{1}{4P\tilde{\mathbf K}^{l/s}\cdot \hat z }\sum_{\eta={\rm I}}^{\rm IV}
\sigma^2_{v^{l/s}}
 + D^{l/s} \bigg ( \frac{\sigma_{|\tilde{\mathbf  K}^{l/s}_D\cdot \hat{z}|}}
{|\tilde{\mathbf  K}^{l/s}_D\cdot \hat{z}|}\bigg )^2\end{eqnarray}
Data for each proton segment are displayed in  Figure \ref{8Ds}, where we
plot values
for the eight individual $D_\eta^{l/s}$.
(The sinusoidal variation is predicted by Equation \ref{Dtransverse}
and also seen in the test data of Figure \ref{fg:ATPDs},
 where the amplitude is 100 times larger.)

The two independent measurements for small angle and large angle
PIN-electron detector pairs
can be combined in a weighted
average.
\begin{equation}
D= \frac { D^s/\sigma^2_{D^s}+D^l/\sigma^2_{D^l}}
{ 1/\sigma^2_{D^s}+1/\sigma^2_{D^l}}
\end{equation}
The full uncertainty includes the uncertainty from
the average neutron beam polarization. 
\begin{equation}
\sigma_D^2=  \Big ( \frac {1}{ 1/\sigma^2_{D_{(s)}}+1/\sigma^2_{D_{(l)}}}
\Big )+ \Big ( D\frac{\sigma_{\bar P}}{\bar P} \Big)^2
\end{equation}
The data are also analyzed by breaking each series
up into individual runs and combining PIN-electron detector pairings in the
same way. The results of these analyses
are consistent. The final result is $(-0.6\pm
1.2)\times 10^{-3}$, where we have assumed the neutron polarization is
$\bar P = (96\pm
2)$\%. This is derived from our measurement of flipping ratio described in
Section
\ref{PolBeam}, with
the assumption that the allowed range ($93\%\le P \le 100\%$) spans
$2\sigma_{\bar P}$.

Finally, we use the scaled results from
the systematic test data
(Figure \ref{fg:ATPDs})
 combined with Monte Carlo simulation studies
to estimate the uncertainty of the Tilt-ATP systematic  effect.
 For the test data, proton detector IV ($\phi_{\rm IV} = 270^{\circ}$)
was not operational.  In calculating $D$ for the test data, only values from
detectors I and III  can therefore be used in Equation \ref{eqcomb}
with a result of ${1\over 2} (D_{\rm I}+D_{\rm III})=(-6.5 \pm 1.4 )\times
10^{-2}$.
Monte Carlo simulations show that for a beam of radius 3 cm, the
$\sin(\phi_{\eta}-\phi_{\sigma})$ behavior of Equation \ref{Dtransverse} is
modified so
that $D_{test} =  {1\over 2}(D_{\rm I}+D_{\rm III})/1.6 = (-4.1 \pm 0.9)
\times 10^{-2}$.
This can be scaled by  $\sin\theta_\sigma$, the ratio of polarization
misalignments  for the data and
test  runs. The individual values of $D_\eta^{l/s}$ shown
in Figure
\ref{8Ds} are used to determine $\theta_{\sigma} = (9 \pm 3) \times
10^{-3}$ radians for the data run. This provides
an upper limit for the uncertainty on the Tilt-ATP systematic effect of
$D$(Tilt ATP)$ < D_{test} \sin\theta_{\sigma}  \le 5.2  \times 10^{-4}$.
Though we use the test  results to estimate this false $D$ effect,
we expect the
cancellation due to beam symmetry to be more complete for the data run
because the
test beam was
intentionally distorted. We therefore consider this upper limit to be a
conservative
estimate of the largest possible false $D$ effect 
\cite{BeamProfFootnote}
The contributions to the statistical and systematic uncertainties
are given in Table \ref{errortable}.

  \begin{table} [h]
\begin{center}
\begin{tabular}{l l}
Sources of uncertainty&Contribution ($\times 10^{-4}$)\\
  \hline
  \hline
Statistics&12 \\
\hline
Tilt ATP&5\\
Twist ATP&$<$1\\
Flux variations&negligible\\
Polarization variations&negligible\\
\end{tabular}
\end{center}
\caption{{Contributions to the uncertainty.}}
\label{errortable}
\end{table}

\section{Summary and Conclusions}\label{concl}

The apparatus used to perform a measurement of the
$D$-coefficient in the beta-decay of polarized neutrons has been
described.  The data using the emiT detector have been analyzed using a
technique that is insensitive to the nonuniform detection efficiency over
the proton
detectors.  The initial run
produced a statistically limited result of
$D= \lbrack -0.6\pm 1.2({\rm stat}) \pm 0.5({\rm syst}) \rbrack \times
10^{-3}$.
This result can be combined with earlier measurements to produce a new
world average for the neutron $D$ coefficient of $-5.5\pm 9.5 \times 10^{-4}$,
which constrains the phase of $g_A/g_V$ to $180.073^\circ \pm 0.12^\circ$.
This represents a 33\%  improvement (95\% C.L.) over limits set by the current world
average, and correspondingly
further constrains standard model extensions with leptoquarks 
\cite{Herc95}. The result
is also
interesting in light of upper limits provided by the
neutron and $^{199}$Hg electric dipole moments on T-odd, P-even
interactions such as left-right symmetric models and exotic fermion models.

A second run is being planned with
strategies to improve the statistical limitations  related to
background
 experienced in the first run.  Our study of systematic
effects presented here shows that the largest is the tilt-ATP effect. The
uncertainty
on this effect can be reduced significantly with more data taken in the
transverse
polarization mode described in Section \ref{sec:FalseD}.   With
the planned improvements in place, it will be feasible to
improve the sensitivity  to $D$ to $3 \times 10^{-4}$ or less.

\section*{Acknowledgments}\label{acks}

We would like to
thank Peter Herczeg for many helpful conversations.  
We are also grateful for the significant contributions of  
Steven Elliott.  Thanks are due to Mel Anaya, Allen Myers, 
Tim Van Wechel, and Doug 
Will for technical support and to Vassilious Bezzerides,
Laura Grout, Kyu Hwang, Christopher Scannell, Christina Scovel, and 
Kyle Sundqvist for their work on the project.
We acknowledge the support of the National Institute of Standards and
Technology, U.S. Department of Commerce, in providing the neutron
facilities and other significant supplies and support. This research was made possible in part
by grants from the U.S. Department of Energy Division of Nuclear
Physics (contract numbers DE-AI05-93ER40784, DE-FG03-97ER41020,
>DE-AC03-76SF00098, and  00SCWE324) and the National Science Foundation.  
 L.J.L. would also like 
to acknowledge support from the National Physical Science Consortium 
and the National Security Agency.


\begin{references}
\bibitem{BrandDisclaimer} Certain trade names and company products are mentioned in
the text or
identified in illustrations in order to adequately specify the experimental
procedure and equipment used. In no case does such identification imply
recommendation or endorsement by National Institute of Standards and
Technology, nor does it imply that the products are necessarily the best
available for the purpose.
\bibitem{Dflasefootnote} Exact expressions that do not require polarization
misalignment
effects to be small are
$P\hat\sigma\cdot \tilde{\mathbf K}_D D_{false}(\Delta \Phi) =
(\Delta \Phi/\Phi)(v_0^{b2:a1} u_0^{b2:a1})$
and
$P\hat\sigma\cdot \tilde{\mathbf K}_D D_{false}(\Delta P) = (\Delta P/ 2P)
(v_0^{b2:a1}
 u_0^{b2:a1}).$
Here $v^{b2:a1}$ is given by equation~\ref{eq:kpd},
$u^{b2:a1}=(1/2)(w^{b2} + w^{a1})$, and $v_0^{b2:a1}$
and $u_0^{b2:a1}$ would be measured with $\Delta\Phi=0$ and $\Delta 
P=0$.
\bibitem{BeamProfFootnote}  The beam profiles
  were mapped at the entrance to the detector by measuring the flux
  through a pinhole as it was scanned in two dimensions across the
  beam.  In both cases the beam profiles are highly symmetric.  Even
  with half of the beam blocked,  mixing in the downstream guide tubes
 gives a beam
  density with a center of mass displaced by only 2 mm 
  from the beam
  axis.  For the data run,  the beam was centered to within 0.6 mm.
\end{references}
\end{document}